# The gravitational phase shift test of general relativity

Jean Paul Mbelek
Service d'Astrophysique, CEA-Saclay, France

The aim of this paper is to study the extra phase shift that general relativity (GR) predicts for a radial light ray propagating in the vicinity of a static spherical symmetric body. It appears that the gravitational phase shift test yields a better sensitivity than the gravitational frequency shift or the excess time delay of the photons. An experiment is proposed for this new test of GR pertaining exclusively to the wave aspect of light. To start with, let us use the Schwarzschild metric to describe the gravitational field of the Earth, thus

$$ds^2 = g_{\mu\nu} dx^\mu dx^\nu = (1 - 2(GM/rc^2))c^2 dt^2 - (1 - 2(GM/rc^2))^{-1} dr^2 - r^2(d\theta^2 + \sin^2\theta\, d\varphi^2), \quad (1)$$

hence the gravitational frequency shift [1, 2] and the equation of motion of light rays (null geodesic $ds^2 = 0$) read respectively

$$\nu = \nu_0 (1 - 2(GM/rc^2))^{-1/2} \quad (2)$$

and

$$g_{\mu\nu} k^\mu k^\nu = (1 - 2(GM/rc^2))(\omega^2/c^2) - (1 - 2(GM/rc^2))^{-1} k_r^2 - r^2(k_\theta^2 + \sin^2\theta\, k_\varphi^2) = 0, \quad (3)$$

where $\nu_0 = \omega_0/2\pi$ and $\nu = \omega/2\pi$ are respectively the frequency of the light ray far from any gravitational source and at radius $r$ from its centre of mass ; $k_r$, $k_\varphi$ and $k_\theta$ are the components of the wave vector in spherical coordinates.

Moreover, the spacetime variation of the phase $d\Phi$ induced by the gravitational potential $V_N(r) = - GM/r$ between the radii $r$ and $r + dr$ is given by

$$d\Phi = g_{\mu\nu} k^\mu dx^\nu = (1 - 2(GM/rc^2))\omega\, dt - (1 - 2(GM/rc^2))^{-1} k_r\, dr - r^2(k_\theta\, d\theta + \sin^2\theta\, k_\varphi\, d\varphi), \quad (4)$$

whereas the spatial variation of the phase, which is the one that is of interest to our purpose here, restricts to

$$d\phi = (1 - 2(GM/rc^2))^{-1} k_r\, dr + r^2(k_\theta\, d\theta + \sin^2\theta\, k_\varphi\, d\varphi). \quad (5)$$

As a first step, let us consider a radial light ray ($k_\theta = k_\varphi = 0$, and $\theta$ and $\varphi$ are fixed) propagating from radius $r_i$ to $r_f$. In that case, relation (5) together with relations (2,3) yield a difference $\Delta\phi$ between the GR phase difference and the corresponding flat spacetime value. This extra phase shift amounts to

$$\Delta\phi = (\omega_0/c) \int [(1 - 2(GM/rc^2))^{-1/2} - 1] dr$$
$$\approx (2\pi/\lambda_0)(GM/c^2) \ln(r_f/r_i), \qquad (6)$$

so that $\Delta\phi < 0$ for $r_f < r_i$ and $\Delta\phi > 0$ for $r_f > r_i$.

As an estimate, let the light source of wavelength $\lambda_0 = c/\nu_0$ be at the surface of the Earth and the receiver at a height h above on the same vertical, namely $r_i = R$ and $r_f = R + h$. Hence relation (6) yields

$$\Delta\phi \approx (2\pi R/\lambda_0)gh/c^2, \qquad (7)$$

where $R \approx 6.4\ 10^6$ m denotes the mean radius of the Earth, $g \approx 9.8$ m s$^{-2}$ is the gravitational acceleration of the Earth and we have assumed h << R.

Using a He-Ne laser whose wavelength is $\lambda_0 = 632.8$ nm, and setting the receiver at the height h = 2.4 m above the laser source, one finds $|\Delta\phi| \approx 1°$. This is more than enough needed for the measurement, since the precision reached in laser interferometry is about 1 arcmin. In comparison, the extra time delay of the photons [3-6] of the same light ray reads

$$\Delta t = \int [(1 - 2(GM/rc^2))^{-1} - 1]dr/c \approx 2(GM/c^3) \ln(r_f/r_i) \approx$$
$$2Rgh/c^3 \approx 1.1\ 10^{-17} \text{ s.} \qquad (8)$$

Clearly, the latter result is at present at least an order of magnitude too low to be measured even by the best laboratory atomic clocks.

In figure 1, an experimental set-up is proposed which allows to test the gravitational phase shift given by the general relativistic relation (7). It consists of two fixed mirrors (M1, M2) and a movable one M. Two identical laser sources (S1, S2) and a photodetector PD are used. The whole apparatus may be rotated around the axis (O'O) so that it can be fixed in two basic configurations such that either the plane (S1 O1 O O' O2 S2) is horizontal (horizontal configuration) or it is vertical (vertical configuration). In the first case the path of the light rays (1') and (2') are horizontal, in the second case the path of the light rays (1') and (2') are vertical. In addition, the laser sources (S1), (S2) and the beam splitter BS are tilt controlled in order to ensure the horizontality of the light rays (1), (2) and (3). First, the apparatus is set in the horizontal configuration. Then, the movable mirror is tuned so that the phase difference between the two rays (1) and (3) incident on the photodetector PD is equal to zero and the current I generated as a result of the interference of the light rays (1) and (3) takes the value $I_0$. Thus, when the apparatus is set in the vertical configuration, the gravitational phase shift manifests itself through the detection of a variation $\Delta I = I - I_0 = (I_0/2)(1 - \cos\Delta\phi)$ of the current I due to the gravitational phase shift. In the case the apparatus is not exactly vertical, but tilted with respect to the horizontal by an angle $\alpha$, then the gravitational phase shift reduces by a factor $\sin\alpha$ in the same manner as the height h.

Plotting $|\Delta\phi|$ versus h should yield, by linear interpolation, a slope that equals $2\pi Rg/\lambda_0 c^2 = 0.42°m^{-1}$.

Another way to test the gravitational phase shift would consist to perform the experiment described above in the "Zero-G" Airbus A300 that ESA uses for parabolic flights to generate 25 seconds of free-fall at one time. So, it is during these 25 seconds of weightlessness that the movable mirror M would be quickly tuned such that the phase difference between the two rays (1) and (3) be equal to zero. Thus, after the parabolic flight, the phase difference between the two rays (1) and (3) should be equal to $|\Delta\phi|\sin\alpha$ for a given tilt angle $\alpha$, where $\Delta\phi$ is given by relation (7). Furthermore, one could test at the same time the cancellation of the gravitational frequency shift when the "Zero-G" Airbus A300 is achieving the parabolic flight by comparing the ticks of an atomic clock on board with respect to those of an atomic clock on Earth. To our knowledge, the latter test has not been performed since Vessot et al. experiment [7].

Also, the gravitational phase shift may be tested in an elevator. According to the equivalence principle, one would measure the phase differences $\phi_+ = \phi_0 + (2\pi Rh/\lambda_0 c^2)(g + a)$ and $\phi_- = \phi_0 + (2\pi Rh/\lambda_0 c^2)(g - a)$ when the elevator is accelerated respectively upward and downward with an acceleration of magnitude a. Substracting, one gets $\phi_+ - \phi_- = (4\pi Rh/\lambda_0 c^2)a$.

References

[1] R. V. Pound and G. A. Jr. Rebka, Phys. Rev. Lett. 3, 439 (1959).

[2] R. V. Pound and J. L. Snider, Phys. Rev. Lett. 13, 539 (1964).

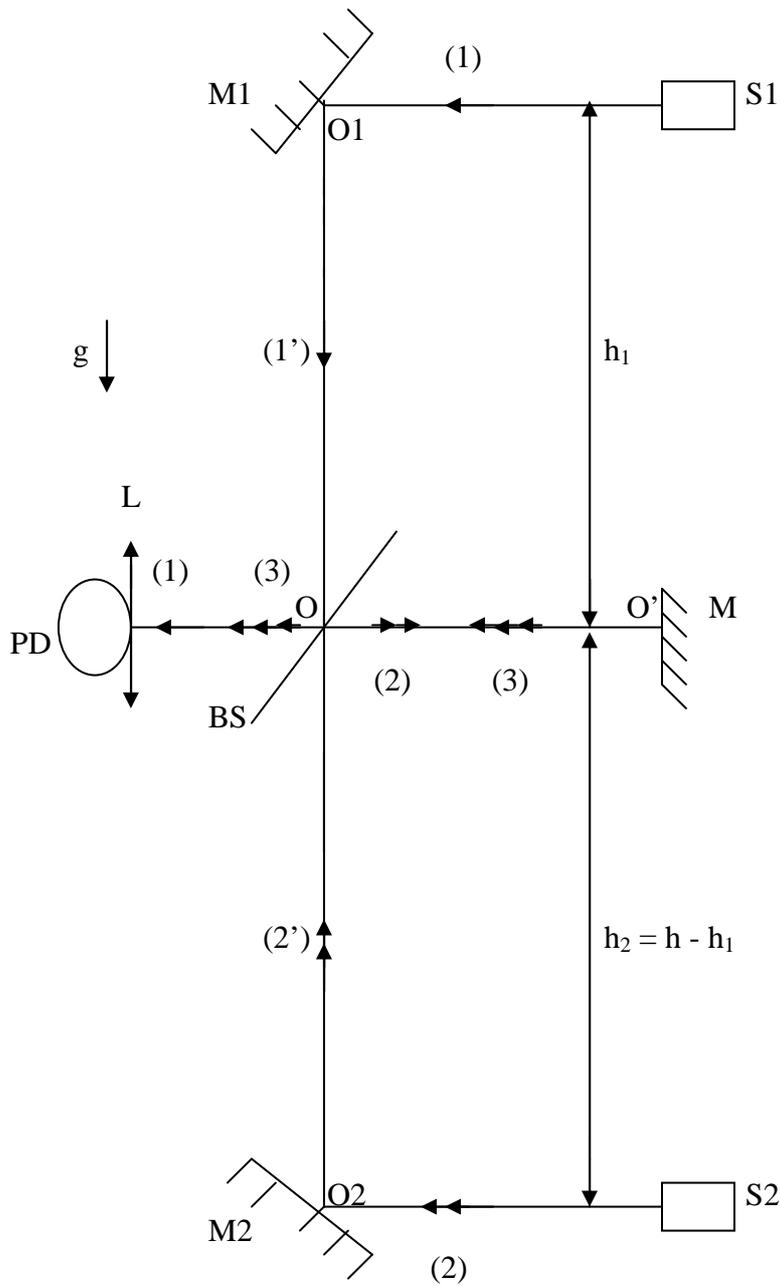

FIG. 1 : Experimental set-up : S1 and S2 laser sources. BS beam splitter. L convergent lens. M1 and M2 fixed mirrors. M movable mirror. PD photodetector.